\documentclass{article}
\usepackage[utf8]{inputenc}
\usepackage{amsthm}
\usepackage{url}
\usepackage{tcolorbox}
\usepackage[affil-it]{authblk}
\usepackage{xcolor}
\usepackage{tikz}

\usepackage{float}
\newfloat{defbox}{tbp}{lob}
\floatname{defbox}{Box}

\usetikzlibrary{arrows.meta,positioning,automata}
\tikzset{state/.style={circle, draw, minimum size=2em}}
\usepackage{amsmath,amsfonts}
\usepackage{booktabs}
\usepackage{array}
\usepackage{tabularx}
\usepackage{microtype}
\usepackage[hidelinks]{hyperref}

\newtheorem{example}{Example}

\newtheorem{remark}{Remark}
\newcommand{\Bodies}{\mathsf{Bodies}}
\newcommand{\Types}{\mathsf{Types}}
\newcommand{\Policy}{\mathsf{Policy}}
\newcommand{\Effect}{\mathsf{Effect}}
\newcommand{\tally}{\mathsf{tally}}

\title{Cardano's Voltaire Governance: \\ Complete Specification and Research Program}
\author[1]{Nimrod Talmon}
\author[2]{Oghenekaro Elem}
\affil[1]{BGU, IOG, \href{mailto:talmonn@bgu.ac.il}{talmonn@bgu.ac.il}}
\affil[2]{Parametrig, \href{mailto:karo@parametrig.com}{karo@parametrig.com}}
\date{}
\begin{document}
\maketitle

\begin{abstract}
Blockchain governance, the set of processes by which decentralized protocols evolve, remains a fundamental challenge in balancing adaptability, security, and stakeholder representation. This technical report analyzes Cardano's Voltaire governance system--the on-chain framework introduced via CIP-1694 and enacted through the Chang hard fork in September 2024--and lays down a corresponding research program.

We make two contributions. First, we provide a complete technical specification of Voltaire's mechanisms, including its three-body architecture, seven governance action types, voting rules, and its constitutional framework; this specification is sufficient for implementation or formal analysis. Second, we establish a research agenda for principled governance optimization, including design of an agent-based simulation platform, analysis of delegation dynamics, optimization of multi-objective parameters, and game-theoretic incentive design; we provide preliminary results, including a \emph{formal governance kernel}: a minimal executable model capturing self-amending governance as a state-transition system and enabling rigorous safety and liveness analysis. 

Our report offers a comprehensive technical overview and invites the research community to advance blockchain governance science through rigorous study of Voltaire as a live, large-scale experiment now managing a treasury valued at approximately \$235 million (1.47B ADA as of early July 2026).%
\end{abstract}

\section{Introduction}\label{sec:intro}

Blockchain governance~\cite{kiayias2022sok} refers to the processes and structures through which decisions are made within blockchain systems. As decentralized platforms evolve beyond purely technical protocols into socio-technical systems, governance becomes central to adaptability, security, sustainability, and participation. Governance mechanisms define how changes are proposed, debated, and implemented, and they vary significantly across platforms.

Digital governance plays a critical role in shaping decentralized ecosystems~\cite{schneider2021modular, vitalikgovernance, morini2025decentralized}. Numerous models have been deployed, particularly in decentralized autonomous organizations (DAOs)~\cite{daogovernance, votingdao}. However, designing effective and scalable governance remains a substantial challenge~\cite{scalingnoteasy, chen2021decentralized}. This report focuses on Cardano's Voltaire governance system as a case study for rigorous analysis.

\paragraph{IT governance in modern information systems}
In context, information systems governance, the structures and processes by which IT decisions are made, has long been recognized as critical to organizational performance. Blockchain systems, as decentralized information infrastructures, inherit this challenge with added complexity: no central authority, pseudo-anonymous stakeholders, and immutable execution. The question ``who decides and how?'' becomes existential when there is no CEO, no board of directors, and no legal jurisdiction.

\subsection{Blockchain Governance: Design Space and Challenges}\label{sec:related-work}

As a specific, high-stakes ecosystem within the IT world, we focus on blockchain governance.
In this sphere, blockchain governance faces a fundamental tension: decentralization requires distributed decision-making, yet coordination costs scale poorly. Early systems like Bitcoin rely on off-chain social consensus; newer systems like Tezos and Polkadot experiment with on-chain mechanisms. The design space spans: who can propose, who votes, what thresholds apply, and how changes enact.

Blockchain governance research has identified several key design dimensions and persistent challenges that structure the field.
\begin{itemize}

\item
\textbf{Governance surfaces:} Systems differ in where decisions are made~\cite{filippi2024report}. \emph{Off-chain governance} (Bitcoin's BIP process, Ethereum's EIP process pre-merge) relies on social consensus: developers propose changes, community debates occur on forums, and miners/validators signal acceptance by running updated software. Legitimacy derives from rough consensus among core developers and economic actors. \emph{On-chain governance} (Tezos~\cite{tezos-governance}, Polkadot~\cite{polkadot-governance,elem2024opengov}) encodes decisions in the protocol: proposals are submitted as transactions, voting occurs cryptographically, and ratified changes execute automatically. Cardano's Voltaire represents a hybrid: CIP proposals are off-chain, but Voltaire's execution is fully on-chain. This report focuses exclusively on Voltaire's on-chain mechanisms.

\item
\textbf{Authority distribution:} Governance authority may concentrate or distribute across stakeholder classes~\cite{kiayias2022sok}. Pure \emph{token-weighted voting} (plutocratic) risks whale dominance. \emph{Validator/miner voting} (infrastructure-weighted) risks centralization in infrastructure pools. \emph{Core developer authority} (technocratic) enables expertise but risks benevolent dictatorship. \emph{Hybrid multi-stakeholder models} (Voltaire's three-body system) aim for checks-and-balances but introduce complexity. No consensus exists on optimal design; tradeoffs remain empirically underexplored.

\item
\textbf{Persistent challenges:} Blockchain governance faces tensions absent in traditional systems~\cite{vitalikgovernance, schneider2021modular}: \emph{immutability vs.\ adaptability} (protocols resist change by design, yet must evolve to fix bugs and add features); \emph{anonymity vs.\ accountability} (pseudonymous participants complicate reputation systems and enforcement); \emph{global coordination} (no geographic boundaries or legal jurisdiction; decisions must align globally distributed actors with heterogeneous incentives); and \emph{economic attacks} (governance becomes a financial asset, where vote buying and bribery are economically rational strategies).

\item
\textbf{Comparative frameworks:} Recent work~\cite{filippi2024report} proposes evaluating governance across five dimensions: organizational profile (legal structures, founding entities), governance areas (monetary policy, protocol upgrades, treasury management), frameworks (decision-making rules), surfaces (on-chain vs. off-chain), and trends (decentralization trajectory). Existing studies~\cite{kiayias2022sok} survey mechanisms but lack empirical benchmarks. This gap motivates our dual contribution: a definitive Voltaire specification enabling formal modeling, and a research agenda for simulation-driven optimization.

\item
\textbf{Positioning Voltaire:} Cardano's governance is distinguished by high parameterizability ($\sim$30 tunable thresholds), formal specification (CIP-1694), and a multi-body architecture balancing economic stake (DReps), infrastructure operation (SPOs), and constitutional review (CC). These characteristics make Voltaire particularly amenable to systematic study. Systems with hardcoded constants or informal processes are harder to analyze this way.

\end{itemize}

\subsection{Cardano's Voltaire: A Case Study}\label{sec:voltaire-case-study}

Cardano represents a research-driven, systematically engineered approach to decentralized infrastructure, distinguishing it from many blockchain projects that prioritize rapid deployment over rigorous design. From its inception, Cardano has emphasized peer-reviewed research, formal methods, and modular architecture. Those principles extend to its governance system. This research orientation makes Voltaire particularly valuable as a subject for technical analysis and scientific study.
In particular, the Voltaire governance system, introduced via CIP-1694 and ratified through the Chang hard fork (September 2024), embodies several design characteristics that facilitate systematic investigation:
\begin{itemize}

\item \textbf{Formal specification:} CIP-1694 provides detailed ledger rules and state transition semantics, making principled analysis tractable.

\item \textbf{High parameterizability:} Voltaire exposes $\sim$30 governable parameters (thresholds, deposits, quorums, etc.), enabling controlled experimentation on how parameter choices affect attack costs, participation, and proposal latency. Unlike hardcoded systems, this allows empirical optimization.

\item \textbf{Multi-body architecture:} The three-body model (DReps, SPOs, CC) creates a complex design space reflecting tradeoffs between efficiency, security, and representation. Studying it illuminates fundamental questions: How do multi-stakeholder systems balance competing interests? When do checks-and-balances improve security vs. create gridlock?

\end{itemize}

In summary, Voltaire's combination of research rigor, formal specification, and parameterizability makes it an ideal testbed for governance science. It connects theory, such as formal models and game-theoretic analysis, with live experimentation on a ${\sim}\$6$B blockchain. The following sections build on that foundation to present both a definitive technical reference and a research agenda for principled governance analysis.

\subsection{Paper Objectives and Scope}

This report serves two main purposes. First, we provide a complete technical specification of Voltaire's mechanisms, including its three-body architecture (Delegated Representatives, Stake Pool Operators, Constitutional Committee), seven governance action types, voting rules, and constitutional framework. This specification is sufficient for implementation or analysis without external dependencies.
Second, we establish a research agenda for principled governance analysis and optimization, including formal verification, agent-based simulation, game-theoretic equilibrium analysis, and mechanism design. We discuss preliminary work, including a \emph{formal governance kernel}--a minimal executable model capturing self-amending governance as a state-transition system--and identify key challenges in complexity, centralization risk, and participation; concretely, we state open questions on safety properties, threshold calibration, and incentive alignment.

We focus on on-chain mechanisms (CIP-1694), treating off-chain processes (Cardano Improvement Proposals, social consensus) as context. We do not compare across blockchains extensively; Cardano is the object of study. Beyond presenting governance mechanisms, we aim to lay groundwork for agent-based simulations to evaluate resilience under varying participation levels, strategic behavior (delegation, abstention, bribery), and stake distributions.

\subsection{Paper Structure} 

Section~\ref{sec:background} provides background on Cardano's evolution leading to Voltaire. Section~\ref{sec:specification} presents the complete technical specification, covering the three-body architecture, action types, voting thresholds, and enactment procedures. Section~\ref{sec:researchagenda} introduces the formal governance kernel and Markov-chain interpretation, then outlines the research agenda, including a simulation framework, safety--liveness analysis, delegation dynamics, and parameter optimization. Section~\ref{sec:conclusion} concludes with a call to action.

\section{Background: the Pre-Voltaire Era}\label{sec:background}

Cardano was launched in 2017 as a third-generation blockchain, distinguishing itself through peer-reviewed research and formal methods. Its development is structured around five eras, each adding capabilities:

\begin{enumerate}
    \item \textbf{Byron (2017--2020):} Foundation layer. Federated block production (IOG-controlled). No on-chain governance; protocol changes required software updates coordinated by IOG.

    \item \textbf{Shelley (2020):} Decentralized block production via Stake Pool Operators. ADA holders began delegating stake to SPOs for rewards. Nakamoto coefficient improved from 1 (Byron) to $\sim\!20$. Still no on-chain governance for protocol upgrades.

    \item \textbf{Goguen / Alonzo (2021):} Smart contract capability via Plutus. Enabled programmable logic but governance remained off-chain. Key design decision: Cardano chose safe, formally verified smart contracts over EVM compatibility.

    \item \textbf{Basho (2022--2024):} Scalability improvements (input endorsers, pipelining, Hydra L2). These changes were enacted through soft forks coordinated by IOG. This meant decision-making was still centralized despite SPO decentralization.

    \item \textbf{Voltaire (2024--present):} Governance era. Defined by CIP-1694, ratified via Chang hard fork (September 2024). Cardano's first comprehensive on-chain governance framework. Introduced DReps, the Constitutional Committee, and the governance action framework. A temporary \emph{Interim Constitution} was adopted at Chang; the permanent Constitution was approved by 95\% of delegates at the Constitutional Convention in Buenos Aires and Nairobi (December 4--6, 2024), put to an on-chain vote on January 30, 2025, reached the DRep approval threshold of 75\% with unanimous Interim Constitutional Committee (ICC) approval, and was enacted on-chain on February 23, 2025~\cite{iog-constitution}.
\end{enumerate}

\paragraph{The CIP process.}
Cardano Improvement Proposals (CIPs) are the off-chain mechanism for proposing technical standards. CIP-1694 (``A First Step Towards On-Chain Governance,'' 2023) is the foundational specification for Voltaire, authored by the Cardano community and IOG after $50+$ global workshops and testnet deployment on SanchoNet. While CIPs govern what the community proposes, Voltaire's on-chain mechanisms handle ratification and execution. This report focuses exclusively on the latter.

\paragraph{Motivation for Voltaire.}
Before Voltaire, Cardano's governance was \emph{de facto} controlled by IOG, the Cardano Foundation, and EMURGO, the three founding entities. While technically capable, this arrangement posed sustainability risks: the community lacked formal mechanisms to participate in protocol evolution, treasury allocation, or constitutional review. Voltaire addresses this by progressively decentralizing decision-making to ADA holders and their elected representatives, with the founding entities transitioning to participants rather than controllers.

\section{Voltaire Governance: Complete Specification}\label{sec:specification}

Cardano's Voltaire governance was defined by CIP-1694~\cite{cip1694} (``A First Step Towards On-Chain Governance''), drafted in 2023 through a collaborative process involving IOG researchers, community representatives, and SPO delegates. The proposal underwent extensive public review via 50+ global workshops, multiple draft revisions, and testnet deployment (SanchoNet), before being ratified via the Chang hard fork in September 2024. 
The system initially operated under an \emph{Interim Constitution} following 
the Chang hard fork. After a global consultation process involving 63 workshops across 52 countries and a Constitutional Convention held simultaneously in Buenos Aires and Nairobi (December 4--6, 2024), the Final Constitution was put to an on-chain vote on January 30, 2025, reached the DRep approval threshold of 75\% with unanimous ICC approval, and was enacted on-chain on February 23, 2025~\cite{iog-constitution}.

CIP-1694 established three governance bodies (Delegated Representatives, Stake Pool Operators, Constitutional Committee), seven action types (no-confidence, committee updates, constitutional amendments, hard forks, parameter changes, treasury withdrawals, info actions), and voting thresholds. The separately ratified Cardano Constitution provides normative principles and guardrails; the Constitutional Committee ensures proposals comply with constitutional constraints. This design reflects Cardano's research-driven philosophy: formal specification enabling verification, high parameterizability enabling optimization, and explicit iteration (``first step'') enabling continuous improvement.

This section specifies the complete on-chain mechanisms. Our goal is sufficiency: a reader should be able to understand, implement, or formally model Voltaire without external references beyond the cited CIP-1694~\cite{cip1694} and Constitution~\cite{cardano-constitution} documents.

\begin{remark}[Scope: on-chain governance only]
Cardano Improvement Proposals (CIPs) are off-chain, community-driven documents for proposing technical standards, protocol enhancements, and governance procedures. CIPs follow a GitHub-based review process but are not enforced by the ledger. While CIP-1694 \emph{defines} Voltaire's mechanisms, the CIP process itself lies outside our formal model. This report focuses exclusively on the on-chain governance system: how proposals are submitted as transactions, how votes are tallied cryptographically, and how ratified actions execute automatically.
\end{remark}

The governance system is supported by a highly parameterized infrastructure, including technical, network, economic, and governance parameters. This design enables fine-grained, on-chain control over core protocol behaviors and supports automated enforcement of governance outcomes~\cite{morini2025decentralized}.

\subsection{Architecture Overview and the Constitution}\label{sec:architecture}

Voltaire implements a \emph{multi-body governance system} with constitutional constraints. At its core are:

\begin{enumerate}
    \item \textbf{Governance actors}: Four classes of participants with distinct roles.
    \item \textbf{Governance actions}: Formal proposals that modify the protocol or allocate resources.
    \item \textbf{Voting mechanisms}: Rules determining how actions are evaluated and ratified.
    \item \textbf{Constitutional framework}: Normative and procedural constraints on actors, actions, and mechanisms.
\end{enumerate}

The system operates in \emph{epochs}, fixed time periods lasting 5 days under the current network configuration. During each epoch, stake distribution and governance state remain stable. Governance actions are submitted, voted on, and, if ratified, enacted on epoch boundaries. This epochal structure simplifies stake calculation and ensures deterministic voting outcomes.

\subsubsection{Three-body model (DReps, SPOs, CC)}\label{sec:three-body}

Cardano's on-chain governance is structured around four principal entities, three of which hold voting power:
\begin{itemize}

\item
\textbf{ADA Holders} are the sovereign actors in the system. They possess the right to vote on governance matters by self-registering as a Delegated Representative (DRep) or delegating to such per stake credential. (Multi-DRep delegation, allowing weighted splitting of voting power across multiple DReps, has been proposed~\cite{cip-x-multidrep} but is not yet part of the protocol.) 

\item
\textbf{Delegated Representatives (DReps)} are registered voting entities who receive voting power from ADA holders. DReps vote on most categories of governance actions and constitute the primary decision-making layer in the governance process. Any ADA holder can register as a DRep by submitting an on-chain registration certificate (requires 500 ADA deposit). Delegation is non-custodial (delegators retain full control of their ADA), revocable (redelegation takes effect next epoch), and stake-preserving (staking rewards are unaffected by governance delegation).

\item
\textbf{Stake Pool Operators (SPOs)} are responsible for block production and also participate in governance by voting on specific action types, such as protocol version updates (as they are responsible for enacting such actions). SPOs vote on hard fork initiations, protocol parameter changes affecting network performance, and constitutional amendments. Their votes are weighted by delegated stake, and hard forks require both on-chain approval and node-level endorsement (SPOs must actually upgrade their software).

\item
\textbf{The Constitutional Committee (CC)} is a fixed-term, multi-member body tasked with verifying the constitutional validity of certain governance actions, based on the Cardano Constitution. The Interim CC established at the Chang hard fork consists of 7 members~\cite{cf-governance-2025}: four appointed (IOG, Cardano Foundation, EMURGO, Intersect) and three elected by the community (Cardano Atlantic Council, Eastern Cardano Council, Cardano Japan), serving a 73-epoch ($\approx$1 year) term. The committee size and composition are governable via ``Update CC'' actions. CC votes are one-member-one-vote (not stake-weighted), requiring $> 67\%$ approval ($\geq 5$ of 7 under the current Interim CC). The CC can be removed via a no-confidence motion and is accountable to the community through this mechanism.

\end{itemize}

\subsubsection{Constitutional framework}\label{sec:constitutional-framework}

The Cardano Constitution serves as the foundational document for the governance system introduced under CIP-1694 and ratified by the Chang hard fork. Unlike static protocol rules, the Constitution is conceived as a \emph{living document}. It is meant to guide decentralized decision-making while adapting to the community's needs over time. It establishes the principles, values, and constraints that govern the legitimacy of on-chain actions and defines the normative framework within which actors such as Delegated Representatives and the Constitutional Committee must operate.

Indeed, Cardano's approach reflects a broader trend in blockchain governance toward formalization and constitutionalism. The Constitution functions as a normative anchor that links social consensus to protocol-level governance, serving as a reference point for validating on-chain actions. While not executable code, it delineates the boundaries of legitimate authority, defines actor responsibilities, and embodies the community's shared principles~\cite{filippi2024report}.

The Cardano Constitution can be accessed online.\footnote{\url{https://constitution.gov.tools/en/constitution}} It contains both the description of abstract principles as well as more concrete topics such as the supermajority needed to amend the constitution itself. The CC is expected to act accordingly. 
The Constitution is stored as an on-chain hash (SHA-256 digest) pointing to an off-chain document. When amended, a ``New Constitution'' governance action is submitted with a new hash, requiring 67\% DRep + 51\% SPO + CC approval. This design balances immutability (changing the Constitution is hard) with adaptability (not impossible).

The Constitution includes:

\begin{itemize}
\item \textbf{Preamble and overview:} The Constitution begins with a declaration of intent that defines decentralized, on-chain governance for the Cardano blockchain. ADA holders play a central role in proposing and deciding on changes to the system, as the sovereign actors.

\item \textbf{Guiding principles (tenets):} The constitution articulates core tenets: no censorship or transaction delays; reasonable and predictable fees; open access for developers; fair compensation and user control over data and funds; long-term sustainability and fairness; and maximum ADA supply fixed at 45 billion, for economic stability.

\item \textbf{Governance structure:} The constitution describes the governance structure (consisting of ADA holders, DReps, SPOs, and the CC) as described in Section~\ref{sec:three-body}.

\item \textbf{Governance process:} The constitution describes the governance process (consisting of on-chain mechanisms for proposing and voting on proposals) with different supermajorities needed (e.g., constitution amendments requiring 67\% DRep approval of the active voting stake; and some actions that require acceptance both by DReps and by the SPOs).

\item \textbf{Guardrails:} The constitution concludes with the guardrails, which are technical and procedural constraints that limit what governance actions can do in order to ensure security, performance, and economic stability. These include restrictions on changing parameters such as transaction size, block size, fees, and treasury withdrawals. Some guardrails are enforced automatically via on-chain scripts (Plutus validators that check proposal parameters), while others rely on off-chain review and benchmarking by the CC.
\end{itemize}

\subsection{Governance Actions}\label{sec:governance-actions}

Cardano's governance system centers on formal \emph{governance actions}, which are proposed on-chain and progress through a structured lifecycle. Governance actions are the fundamental unit of on-chain decision-making. They are structured proposals to modify the protocol, allocate resources, or change governance rules. Actions are submitted as special transactions, voted on by relevant bodies, and, if ratified, enacted automatically.

\subsubsection{Action types and parameters}\label{sec:action-types}

Voltaire defines seven action types:

\begin{enumerate}

\item
\textbf{No-confidence motion:} A motion to express no-confidence in the current Constitutional Committee. If ratified (requires $\geq 67\%$ DRep + $\geq 51\%$ SPO approval), the system enters a state of no-confidence, all CC members are removed, and future proposals requiring CC approval are blocked until a new CC is elected.

\item
\textbf{Update committee and/or threshold and/or terms:} A proposal to modify the membership, signature threshold, or terms of service of the Constitutional Committee. Parameters include new CC members (identified by public key hashes), new approval threshold, and term lengths. If ratified (requires $\geq 67\%$ DRep + $\geq 51\%$ SPO approval), the new CC is installed and the no-confidence state (if active) is cleared.

\item
\textbf{New constitution or guardrails script:} A proposal to adopt a new Constitution or update the Guardrails Script, recorded as on-chain hashes. Parameters include new constitution hash (SHA-256) and optional guardrails script (Plutus validator). Requires $\geq 67\%$ DRep + $\geq 51\%$ SPO + CC approval.

\item
\textbf{Hard-fork initiation:} A proposal to activate a non-backward-compatible upgrade of the protocol; assumes a prior software upgrade has occurred. Parameters specify the new protocol version. Requires $\geq 60\%$ DRep + $\geq 51\%$ SPO + CC approval, plus node-level endorsement (SPOs must upgrade their software; if insufficient stake-weighted SPOs upgrade, the fork does not proceed). The exact readiness threshold is a protocol parameter that has varied across forks: the Chang upgrade (September 2024) required 70\% SPO readiness, while the Plomin upgrade (January 2025) required 85\% of stake pools by stake~\cite{emurgo-plomin}. This is the highest-risk action in the system, so high thresholds and SPO veto provide defense in depth.

\item
\textbf{Protocol parameter changes:} Changes to one or more modifiable protocol parameters, excluding upgrades involving major version changes (i.e., hard forks). Cardano has $\sim$30 governable parameters divided into network (block size, transaction limits), economic (fees, rewards), technical (script execution costs), and governance (thresholds, deposits) groups. Network group changes require $\geq 67\%$ DRep + $\geq 51\%$ SPO + CC approval; other groups require $\geq 67\%$ DRep + CC approval (SPOs do not vote). Automated guardrails enforce constitutional bounds on parameter ranges (e.g., the current \texttt{maxBlockBodySize} is 90,112 bytes $\approx$ 90 KB).

\item
\textbf{Treasury withdrawals:} Proposals to withdraw funds from the on-chain treasury. Parameters specify amount (in lovelaces; 1 ADA = $10^6$ lovelaces), recipient address, and justification metadata. Requires $\geq 67\%$ DRep + CC approval (SPOs do not vote). Withdrawals are constrained by the Net Change Limit (NCL), a constitutional guardrail that caps the total ADA withdrawable from the treasury within a community-defined period~\cite{intersect-ncl}. The 2025 NCL was set at 350M ADA annually (of which 347M, or 99.1\%, had been utilized by March 2026~\cite{cgov-dashboard}). As of July 2026, the Treasury holds $\sim$1.47B ADA ($\sim$\$235M)~\cite{cardano-treasury-explorer}.

\item
\textbf{Info actions:} Non-binding signals or informational polls. Parameters include arbitrary metadata (e.g., sentiment surveys). Typically DReps vote with no threshold (pure signaling). No ledger state changes; results recorded on-chain for historical reference.

\end{enumerate}

\subsubsection{Submission and anchoring}\label{sec:submission}

A governance action is submitted on-chain by an eligible proposer via a special transaction containing:
\begin{enumerate}
    \item \textbf{Action type}: Enum specifying which of the 7 types.
    \item \textbf{Parameters}: Type-specific data (e.g., new CC members, treasury amount).
    \item \textbf{Metadata anchor}: Hash + URL pointing to off-chain justification (e.g., detailed proposal document, budget breakdown).

    \begin{remark}
    As on-chain storage is expensive, full proposal text lives off-chain (e.g., IPFS, web servers). The on-chain transaction includes a content-addressed hash (SHA-256 or IPFS CID) and URL, ensuring integrity (hash verification prevents tampering) and accessibility (voters retrieve full context).
    \end{remark}
    
    \item \textbf{Deposit}: 100,000 ADA locked until action resolves ($\sim$\$16,000 as of July 2026).

    \begin{remark}
    The deposits prevent spam: submitting frivolous proposals costs real money. If ratified or expired (fails to reach threshold after several epochs), the deposit is returned.
    \end{remark}
    
\end{enumerate}

Crucially, multiple actions can be active simultaneously. To prevent conflicts (e.g., two treasury withdrawals draining 90\% each), proposals include an optional reference to a prior action and conflict resolution: if two actions of the same type both ratify, the one submitted first takes precedence.

\subsubsection{Lifecycle and state transitions}\label{sec:lifecycle}

Each action proceeds through the following stages:

\begin{enumerate}
    \item \textbf{Submitted/Proposal}: A governance action is submitted on-chain, deposit locked. At the next epoch boundary, it becomes active.
    
    \item \textbf{Active/Voting}: Depending on the type of action, at least two of the governance entities (DReps, SPOs, CC) must vote and agree. DReps and SPOs use token-based supermajority; CC uses one-person-one-vote supermajority. Different action types have different thresholds (all parameters are governable). Voting period spans multiple epochs, typically 3--6.
    
    \item \textbf{Ratified/Enactment}: If all applicable conditions are met (thresholds reached, system not in no-confidence unless action is Update CC), the governance action is enacted at the next epoch boundary. Deposit is returned.
    
    \item \textbf{Expired}: After several epochs (default 6) without reaching thresholds, the proposal is nulled and deposit returns.
    
    \item \textbf{Dropped}: Action superseded by conflicting action or governance state change (e.g., no-confidence nullifies pending actions requiring CC approval).
\end{enumerate}

Votes can be cast/changed anytime during the active period. Tallying occurs at epoch boundaries using a snapshot of stake distribution taken at epoch start (prevents double-voting via stake movement). Ratified actions enact at the next epoch boundary, ensuring predictability, atomicity (multiple ratified actions enact simultaneously), and determinism (all nodes compute identical state transitions).

\subsection{Delegated Representatives (DReps)}\label{sec:dreps}

DReps implement liquid democracy: ADA holders delegate voting power to representatives but retain sovereignty via instant, costless redelegation. This design trades direct democracy's cognitive burden for representative democracy's information aggregation, while avoiding entrenchment through continuous accountability.

\subsubsection{Delegation mechanism}\label{sec:drep-registration}

Practically, delegation works as follows:
\begin{itemize}

\item
\textbf{Registration:} Any ADA holder can register as a DRep by locking a 500 ADA deposit and optionally publishing policy metadata (platform, voting history). The deposit prevents Sybil attacks. Creating fake DReps is cheap computationally but expensive economically.

\item
\textbf{Delegation:} ADA holders delegate via an on-chain certificate specifying target DRep. Delegation is non-custodial (delegators retain asset control), revocable (redelegation takes effect next epoch), and stake-preserving (does not affect staking rewards). Critically, delegation is \emph{free}: no transaction fees beyond standard network costs, and no lock-up periods.

\begin{remark}
There are two meta-options: ``Abstain'', where stake is intentionally excluded from thresholds, and ``No Confidence'', where stake is counted as a Yes on no-confidence motions and as a No otherwise. These allow expressive exit: silence (no delegation), abstention, or protest.
\end{remark}

\end{itemize}

Note how the costless redelegation creates a delegation market with low switching costs. DReps compete on reputation (voting history, expertise) rather than lock-in. This incentivizes responsive representation but may favor popularity over principle. See also the problem of delegator-representative commitment described below.

\subsubsection{Voting power calculation}\label{sec:voting-power}

Each DRep's voting power $W'(d)$ equals their own stake plus delegated stake:
\[
W'(d) = W(d) + \sum_{a : D(a) = d} W(a)
\]
where $W(d)$ is DRep $d$'s own stake, $D(a)$ is delegator $a$'s chosen DRep, and $W(a)$ is delegator $a$'s stake.

Approval for proposal $p$ is stake-weighted. Writing $V_p(d) \in \{\text{Yes}, \text{No}, \text{Abstain}\}$ for DRep $d$'s vote:
\[
\mathrm{approval}^{\text{DRep}}(p) = \frac{\sum_{d \in A_D :\, V_p(d) = \text{Yes}} W'(d)}{\sum_{d \in A_D :\, V_p(d) \neq \text{Abstain}} W'(d)}
\]
Abstain stake is excluded from both numerator and denominator; active DReps who do not vote are counted as No. Only \emph{active} stake (delegated to registered DReps) counts; dormant or undelegated stake is excluded.

\begin{remark}
Voting power uses epoch-start snapshots to prevent double-voting (moving stake mid-epoch). This introduces a temporal lag: delegation changes take 5 days to affect voting power.
\end{remark}

Note that, as of March 2026, approximately 5.8B ADA ($\sim$15.7\% of the $\sim$37B circulating supply) is actively delegated for governance~\cite{cgov-dashboard}. Empirical data reveals striking concentration: the top 10 DReps control 48.7\% of active voting power (2.8B ADA), the top 20 control 63.6\% (3.7B ADA), and the top 50 control 84.8\% (4.9B ADA)~\cite{cgov-dashboard}. Governance Space analytics independently confirm that 64.8\% of voting power is concentrated in the top 20 DReps as of Epoch 618~\cite{governance-space}. Network effects favor incumbents: high-stake DReps gain visibility, attract delegations, and gain more stake. Without active counter-mechanisms such as reputation systems or delegation diversity incentives, concentration may intensify.

\subsubsection{Delegation incentives}\label{sec:drep-incentives}

DReps receive \emph{no direct compensation}; thus, participation is motivated by:
\begin{itemize}
    \item \textbf{Influence}: Control over protocol evolution and treasury allocation.
    \item \textbf{Reputation}: Attracting delegations signals community trust.
    \item \textbf{Indirect benefits}: Ecosystem participants (developers, SPOs, businesses) benefit from good governance.
\end{itemize}

This volunteer model favors altruistic or aligned actors but may deter high-quality representatives who cannot afford unpaid labor. It also creates adverse selection: only those with external incentives (e.g., employees or large ADA holders) participate actively. Voting remains subjective. There are no objective ``bad votes,'' and there are no penalties or slashing in place. Accountability operates via \emph{voice} (delegators complain) and \emph{exit} (redelegation). This market-based accountability assumes delegators monitor DRep behavior, which may not hold if information costs are high.

To have some responsiveness, DReps must vote at least once per 20 epochs (a governable parameter; currently, $\approx$ 100 days) to remain active. Inactive DReps' delegated stake becomes dormant (excluded from thresholds). This prevents zombie representatives while tolerating intermittent participation.

\subsection{Stake Pool Operators (SPOs)}\label{sec:spos}

SPOs are the infrastructure backbone: they produce blocks, validate transactions, and maintain network liveness. In governance, SPOs vote on actions affecting their operational responsibilities. This creates a second legitimacy source distinct from economic stake (DReps): \emph{operational feasibility}. A proposal may be economically popular but technically infeasible. SPO votes filter for practicality.

\subsubsection{Role and voting rights}\label{sec:spo-role}

SPOs vote on three action types:
\begin{enumerate}

\item
\textbf{Hard fork initiation:} Protocol upgrades requiring node software changes. SPOs vote on-chain ($\geq 51\%$ approval required) and must upgrade nodes off-chain. If insufficient stake-weighted SPOs upgrade their nodes, the scheduled upgrade does not proceed. This dual mechanism prevents forced upgrades: even if DReps approve, SPOs can veto by not upgrading.

\item
\textbf{Protocol parameter changes (network group):} Parameters affecting network performance (block size, transaction limits, script memory/CPU costs). SPOs vote because they bear operational costs: larger blocks require more bandwidth/storage; faster block times require lower latency. DReps alone might optimize for user experience (bigger blocks $\to$ cheaper fees) without considering SPO constraints (bigger blocks $\to$ centralization pressure as only well-resourced operators can handle load).

\item
\textbf{Constitutional changes:} Major governance rule modifications. SPOs vote to ensure meta-governance changes maintain system integrity. Their stake-weighted vote provides a check against capture: if DReps/CC collude to entrench power, SPOs can block.

\end{enumerate}

Regarding their voting power, SPO votes are weighted by delegated stake, that is, the total ADA delegated to their pool for block production. This aligns voting power with skin in the game: pools with more stake have more to lose from bad decisions. However, it also concentrates power. As of 2025, $\sim$20 pools control 51\% of stake (Nakamoto coefficient $\approx 20$), creating oligopolistic tendencies.

\begin{remark}
Effectively, SPOs embody \emph{checks-and-balances} between economic and operational interests. DReps maximize economic value (low fees, high throughput); SPOs maximize operational sustainability (feasible parameters, stable protocol). This tension is productive: pure token-voting (DReps alone) risks technical debt; pure validator-voting (SPOs alone) risks stagnation. The hybrid model trades efficiency (slower decisions, more veto points) for security (multiple stakeholder sign-off).
This, however, also creates \emph{veto-player dynamics}: any sufficiently coordinated SPO coalition ($> 49\%$ stake) can block proposals indefinitely.
\end{remark}

\subsubsection{Hard fork initiation protocol}\label{sec:hard-fork-protocol}

Hard forks require two-stage approval: on-chain voting (governance intent) and node-level endorsement (operational feasibility):
\begin{itemize}

\item
\textbf{Stage 1:} Hard fork proposal specifies new protocol version. Requires DRep ($\geq 60\%$), SPO ($\geq 51\%$), and CC approval. If ratified, fork is scheduled for future epoch.

\item
\textbf{Stage 2:} SPOs must upgrade node software before activation. If insufficient SPOs upgrade (exact threshold is a protocol parameter), scheduled upgrade does not proceed and the chain continues on the current version.

\end{itemize}

The design rationale of this prevents governance from approving technically broken upgrades. On-chain votes express \emph{what the community wants}; node upgrades verify \emph{what operators can execute}. Without stage 2, token-weighted voting could force infeasible changes.
This creates a liveness vulnerability. Minority SPO coalitions can block upgrades indefinitely. The design prioritizes safety (no forced upgrades) over agility (slow evolution).

\subsection{Constitutional Committee (CC)}\label{sec:cc}

The CC is an elected, term-limited body acting as constitutional guardians. The Interim CC established at the Chang hard fork has 7 members~\cite{cf-governance-2025}; the committee size is a governable parameter (\texttt{committeeMinSize}) that can be changed via ``Update CC'' governance actions. Unlike DReps (economic stake) and SPOs (operational infrastructure), the CC represents \emph{constitutional legitimacy}: it ensures that proposals comply with foundational principles regardless of majority support. This implements \emph{judicial review} for blockchain governance: pure token-voting risks tyranny of the majority, and the CC provides a normative check.

\subsubsection{CC operation}\label{sec:cc-composition}

CC members are elected via ``Update CC'' governance actions requiring $\geq 67\%$ DRep + $\geq 51\%$ SPO approval. Proposals specify member identities (public key hashes), term lengths (in epochs), and approval threshold ($\theta^{\text{CC}}$, under the Interim CC, $\theta^{\text{CC}} > 67\%$). Members serve fixed terms but can be re-elected.

Operationally, CC votes are \emph{one-member-one-vote} (not stake-weighted). For proposal $p$, the CC approves if:
\[
\frac{|\{c \in A_C : V_p(c) = 1\}|}{|A_C|} \geq \theta^{\text{CC}}
\]
where $V_p(c) \in \{0,1\}$ is the vote of the member $c$. Under the Interim CC, $|A_C| = 7$ and $\theta^{\text{CC}} > 0.67$, so $\geq 5$ members must approve (see, e.g., the ICC vote on the Plomin hard fork~\cite{emurgo-plomin}).

The community can remove the CC via a no-confidence motion ($\geq 67\%$ DRep + $\geq 51\%$ SPO). If ratified, all CC members are removed, the system enters a ``state of no-confidence,'' and proposals requiring CC approval are blocked until a new CC is elected. This is the \emph{nuclear option}, a collective firing mechanism preventing CC capture or incompetence.

\begin{remark}
The CC's one-member-one-vote structure deliberately decouples voting power from wealth. This avoids plutocratic capture: a billionaire cannot buy CC votes by accumulating ADA. However, it creates a principal-agent problem: CC members (agents) are elected by DReps/SPOs (principals) but may not faithfully represent community preferences once in office. The no-confidence mechanism provides accountability, but firing the entire CC is costly (governance paralysis until replacement).
\end{remark}

\subsubsection{Constitutional guardrails}\label{sec:guardrails}

The CC approves four action types: constitutional amendments (including guardrails script updates), hard forks, protocol parameter changes, and treasury withdrawals. Info actions, no-confidence motions, and CC updates do not require CC approval (see Table~\ref{tab:thresholds}).
Guardrails are constraints enforcing constitutional principles:

\begin{itemize}

\item
\textbf{Automated Guardrails:} These are on-chain scripts validating proposals. Example: treasury withdrawals are constrained by the Net Change Limit (NCL), a community-ratified cap on total withdrawals within a defined period (the 2025 NCL was 350M ADA annually~\cite{intersect-ncl}). If a withdrawal would cause cumulative withdrawals to exceed the NCL, the proposal transaction is invalid and rejected before voting begins. These are \emph{hard constraints}: cryptographically enforced, with no human discretion.

\item
\textbf{Off-Chain Guardrails:} These are human-evaluated constraints requiring judgment. Example: ``hard fork proposals must include security audit and benchmarking results.'' The CC reviews evidence and votes accordingly. These are \emph{soft constraints}: rely on CC integrity and expertise.

\end{itemize}

The Constitution specifies which parameters have which guardrail types. Automated guardrails are preferable (trustless enforcement) but cannot capture all normative constraints (e.g., ``does this treasury proposal align with ecosystem goals?'' requires subjective judgment).

\subsection{Voting Mechanisms}\label{sec:voting-mechanisms}

Voltaire's voting rules encode tradeoffs between agility (easy to pass proposals) and security (hard to pass malicious proposals). Different action types have different thresholds, reflecting their risk profiles.

\subsubsection{Approval thresholds (by action type)}\label{sec:thresholds}

Table~\ref{tab:thresholds} specifies approval thresholds for each action type and voter class. An action is ratified only if it meets all applicable thresholds.

\begin{table}[t]
\centering
\small
\renewcommand{\arraystretch}{1.15}
\caption{Approval thresholds by governance action. ``--'' indicates that the voter class does not participate. Percentages refer to \emph{active voting stake} (DReps/SPOs) or \emph{active committee members} (CC). All thresholds are governable protocol parameters.}
\label{tab:thresholds}
\vspace{0.5em}
\begin{tabular}{lccc}
\toprule
\textbf{Action type} & \textbf{DReps} & \textbf{SPOs} & \textbf{CC} \\
\midrule
No-confidence                   & $\ge 67\%$ & $\ge 51\%$ & -- \\
Update constitutional committee & $\ge 67\%$ & $\ge 51\%$ & -- \\
New constitution                & $\ge 67\%$ & $\ge 51\%$ & $> 67\%$ \\
Hard fork                       & $\ge 60\%$ & $\ge 51\%$ & $> 67\%$ \\
Protocol parameters (network)   & $\ge 67\%$ & $\ge 51\%$ & $> 67\%$ \\
Protocol parameters (other)     & $\ge 67\%$ & --         & $> 67\%$ \\
Treasury withdrawal             & $\ge 67\%$ & --         & $> 67\%$ \\
Information action              & varies     & --         & -- \\
\bottomrule
\end{tabular}
\end{table}

The design rationale is as follows:
\begin{itemize}
\item \textbf{High-risk actions require higher thresholds}: Constitutional changes ($67\%$ DRep) vs. hard forks ($60\%$ DRep). The constitution is normative bedrock; changing it should be harder than protocol upgrades.

\item \textbf{Meta-actions require multi-body approval}: No-confidence and Update CC actions require approval by both DReps and SPOs, preventing unilateral capture. If DReps could fire the CC alone, they could install a rubber-stamp committee.

\item \textbf{SPOs vote only on operational matters}: They approve hard forks (must upgrade nodes) and network parameters (affect performance) but not treasury withdrawals (purely economic). This limits their power to their domain of expertise.

\item \textbf{CC provides constitutional check}: Most actions require CC approval, except those targeting the CC itself (no-confidence, Update CC) to avoid deadlock. Info actions also bypass CC (non-binding signals need no constitutional review).

\end{itemize}

\subsubsection{Quorum requirements}\label{sec:quorum}

Voltaire uses \emph{active stake thresholds} rather than quorum requirements. The distinction:

\begin{itemize}
\item \textbf{Quorum}: Minimum \% of \emph{total stake} that must vote for a decision to be valid. Example: ``$\geq 50\%$ of total ADA must participate.''
\item \textbf{Active stake threshold}: Approval calculated as \% of \emph{active voting stake} (stake delegated to registered DReps/SPOs). Dormant stake is excluded.
\end{itemize}

Voltaire uses the second approach. Approval for proposal $p$ is:
\[
\mathrm{approval}^{\text{DRep}}(p) = \frac{\sum_{d \in A_D :\, V_p(d) = \text{Yes}} W'(d)}{\sum_{d \in A_D :\, V_p(d) \neq \text{Abstain}} W'(d)}
\]

where the denominator is \emph{active, non-abstaining stake only}. If only $\sim$15.7\% of total ADA is actively delegated for governance (as observed in March 2026~\cite{cgov-dashboard}), a proposal needs $\geq 67\% \times 15.7\% \approx 10.5\%$ of total supply to pass.

\begin{remark}
The rationale is to avoid perpetual gridlock (to improve liveness): e.g., if quorum = 50\% of total stake but only 15\% participates, \emph{no proposal can ever pass}. Active stake thresholds allow governance to function with realistic participation.
Furthermore, this may incentivize participation: low participation $\to$ lower effective threshold (easier for active minority to pass proposals) $\to$ pressure on passive holders to delegate and activate their stake.
Finally, this also introduces a safety risk: if participation is very low (e.g., 10\% active), governance becomes vulnerable to capture. A small coalition controlling 7\% of total stake could pass proposals requiring 67\% approval. Mitigations include DRep activity requirements (inactive DReps' stake becomes dormant) and community monitoring.
\end{remark}

\subsection{Treasury Management}\label{sec:treasury}

Cardano's on-chain treasury is a collectively-owned fund ($\sim$1.47B ADA, $\sim$\$235M as of July 2026~\cite{cardano-treasury-explorer}) financing ecosystem development, infrastructure, and public goods. Treasury management exemplifies blockchain governance's core tension: how to allocate pooled resources without central authority.

\subsubsection{Funding proposals}\label{sec:funding}

To request treasury funds, a proposer submits a ``Treasury Withdrawal'' governance action specifying:
\begin{itemize}
    \item \textbf{Amount}: Requested ADA (in lovelaces).
    \item \textbf{Recipient}: On-chain address receiving funds.
    \item \textbf{Justification}: Off-chain metadata (project description, budget breakdown, milestones, team credentials).
\end{itemize}

The proposer must lock a 100,000 ADA deposit (same as other governance actions). The proposal enters the standard lifecycle: submitted $\to$ active $\to$ voted on by DReps and CC $\to$ ratified/expired/dropped.

Treasury withdrawals require $\geq 67\%$ DRep approval and CC approval ($> 67\%$ of committee members). SPOs do not vote because treasury allocation is an economic decision, not an operational one.
The CC's role is critical: they review proposals against constitutional guardrails (automated:  cumulative withdrawal constraints via the NCL; off-chain: project alignment with ecosystem goals, budget reasonableness, team accountability). A proposal passing DRep vote but failing CC review is blocked.

\subsubsection{Budget constraints}\label{sec:budget}

The Constitution mandates a \emph{Net Change Limit} (NCL), a community-ratified cap on the total ADA withdrawable from the treasury within a defined period~\cite{intersect-ncl}. The NCL is set via an Info Action voted on by DReps and can be adjusted as priorities evolve:

\[
a_{\text{cumulative}} \leq \text{NCL}
\]
where $a_{\text{cumulative}}$ is the sum of all treasury withdrawals within the NCL period and NCL is the community-ratified cap (the 2025 NCL was set at 350M ADA annually). If a proposed withdrawal would cause cumulative withdrawals to exceed the NCL, the proposal transaction is invalid and rejected before voting begins.
This prevents treasury depletion: even with 67\% DRep + CC approval, a series of withdrawals exceeding the NCL cannot pass.

The NCL balances two concerns:
\begin{itemize}
    \item \textbf{Too high}: Allows treasury depletion (rapid drawdowns in a short period leave insufficient funds for future needs)
    \item \textbf{Too low}: Limits responsiveness (emergency funding or large strategic initiatives may require multiple NCL periods to fund)
\end{itemize}

\begin{remark}
Voltaire lacks formal budgeting mechanisms (e.g., ``allocate 50M ADA/year to core development''). Each proposal is atomic and approved or rejected in isolation. This creates several issues:
\textbf{Unpredictability}: Projects cannot rely on multi-year funding commitments;
\textbf{Coordination failure}: Multiple proposals for similar projects may pass or all fail, leading to duplication or gaps; \textbf{Short-termism}: Voters favor immediate-impact proposals over long-term investments
\end{remark}

\begin{remark}
Cardano operates a parallel off-chain treasury system, Project Catalyst, which allocates smaller grants (\$1K--100K ADA) via community voting. Catalyst uses a different model: quadratic voting, reputation-weighted curation, and specialized funding rounds. Voltaire's on-chain treasury is intended for larger, strategic initiatives (core infrastructure, major partnerships).
\end{remark}

\subsection{Update and Enactment Procedures}\label{sec:enactment}

Once a governance action is ratified, it must execute by updating ledger state, protocol parameters, or triggering off-chain processes such as hard fork node upgrades. Enactment procedures ensure these updates occur deterministically and safely.

All ratified actions enact at \emph{epoch boundaries}, fixed points every 5 days when stake snapshots update and rewards are distributed. No mid-epoch enactment occurs. This design ensures:

\begin{itemize}
\item \textbf{Predictability}: All network participants (SPOs, users, exchanges) know exactly when changes take effect. No surprise parameter shifts mid-operation.

\item \textbf{Atomicity}: Multiple ratified actions enact simultaneously in a single state transition. This avoids partial states (e.g., ``treasury withdrew but parameter didn't update'').

\item \textbf{Determinism}: All nodes compute identical state transitions using the same epoch boundary snapshot. No race conditions or node-specific timing issues.
\end{itemize}

Different action types have different enactment semantics:
\begin{itemize}
\item \textbf{No-confidence}: Sets $\sigma[\texttt{no\_confidence}] := \texttt{true}$ and removes all CC members ($A_C := \emptyset$). Proposals requiring CC approval are blocked until a new CC is elected.

\item \textbf{Update CC}: Installs new CC members ($A_C := A_C'$), updates threshold ($\theta^{\text{CC}} := \theta'$), and clears no-confidence state ($\sigma[\texttt{no\_confidence}] := \texttt{false}$).

\item \textbf{New constitution}: Updates constitution hash ($\sigma[\texttt{constitution\_hash}] := h'$) and optionally replaces guardrails script.

\item \textbf{Hard fork}: Schedules protocol version upgrade. It does \emph{not} immediately change protocol. It waits for SPO node-level endorsement (Stage 2 of hard fork protocol). If a sufficient fraction of stake-weighted SPOs runs the new version by the scheduled epoch (the exact threshold is a protocol parameter; e.g., 85\% for the Plomin upgrade~\cite{emurgo-plomin}), the fork activates. Otherwise, the action is nullified.

\item \textbf{Protocol parameter changes}: Sets $\sigma[\texttt{paramName}] := v'$ for each changed parameter. Changes take effect at the next epoch boundary. For example, if max block size increases, new blocks use the new limit.

\item \textbf{Treasury withdrawals}: Transfers funds ($\sigma[\texttt{treasury}] := \sigma[\texttt{treasury}] - a$) to recipient address. Transfer executes as a special transaction at epoch boundary.

\item \textbf{Info actions}: No state changes. Proposal metadata is recorded on-chain for historical reference.
\end{itemize}

If multiple actions of the \emph{same type} ratify in the same epoch, only the first (by submission order) enacts. Others are nullified. Example: two ``Update CC'' proposals both pass, the first proposal's CC is installed, and the second is dropped.
If actions of \emph{different types} ratify, all enact simultaneously. Example: Epoch 500 sees a constitution change, parameter update, and treasury withdrawal all enact together.
(This ordering rule prevents double-spending, e.g., two treasury withdrawals draining 90\% each cannot both execute, while allowing independent actions to compose.)

Enactment is fail-safe: if an action cannot execute (e.g., treasury withdrawal amount exceeds balance due to concurrent withdrawals), the action is \emph{dropped} and the deposit is returned. The chain continues operating with no halting or rollback.
However, this creates a liveness risk: if guardrails are poorly calibrated or state evolves unexpectedly (e.g., treasury depletes faster than anticipated), legitimate ratified proposals may fail to enact. Strict guardrail design is critical; see Section~\ref{sec:guardrails}.

\section{Research Program}\label{sec:researchagenda}

With Voltaire's technical specification established, a natural question arises:
how can we make it work \emph{well}? This section lays out our research agenda
for blockchain governance, building up in three stages before arriving at
concrete research directions. First, we discuss the conceptual shift from
specification to optimization--having specified how Voltaire works, the
challenge now is to study and improve its governance mechanisms
(Section~\ref{section:optimizing governance}). Next, we identify several concrete
challenges facing Voltaire that directly motivate the research directions to
follow (Section~\ref{section:voltaire problems}). We then introduce our main
methodological vehicle: a four-layer architecture for blockchain governance
simulation modeling (Section~\ref{section:gg}). With these pieces in place,
we present concrete research directions in
Section~\ref{section:research directions}.

\subsection{From Specification to Optimization}\label{section:optimizing governance}

\paragraph{What is governance solving?}

To speak of optimizing governance, we have to understand what is governance solving. Decentralized governance must balance three broad, competing objectives.
The first is \textit{legitimacy and social welfare}: decisions should reflect
authentic stakeholder preferences and maximize collective benefit, demanding both
representational accuracy and efficiency--minimizing cognitive burden,
transaction costs, and decision latency.
The second is \textit{security and safety}: the system must resist capture and
manipulation, whether economic (vote buying, bribery, stake concentration) or
social (misinformation, coordination attacks), while remaining stable, coherent,
and predictable.
The third is \textit{adaptability and liveness}: governance must evolve to
address new challenges--protocol bugs, economic shifts, emerging threats,
changing community needs--without ossifying into irrelevance.

These objectives conflict in fundamental ways. High approval thresholds (e.g.,
Voltaire's 67\% DRep requirement for constitutional changes) enhance security
against minority capture but create veto power and potential gridlock, harming
adaptability. Broad participation strengthens legitimacy but imposes attention
costs and slows decision-making. Frequent amendment enables responsiveness but
risks instability. The central challenge is not to ``optimize'' governance in any
absolute sense, but to \emph{characterize these tradeoffs} and identify
configurations that achieve acceptable performance across multiple dimensions.

\paragraph{From specification to optimization.}

The preceding sections provided a complete technical specification of Voltaire--a
sophisticated, carefully engineered attempt to navigate precisely these tensions,
reflecting years of research, community consultation, and iterative refinement.
But specifying a governance system is not the same as understanding whether it
works well, or \emph{how} well. Having established \emph{what} Voltaire does, our
research agenda turns to studying how effectively it does it, and how it might be
improved.

Our approach is to model governance as an executable, self-amending multiagent
system amenable to formal analysis, agent-based simulation, empirical
calibration, and multi-objective optimization. While we use Voltaire as the
primary case study, the methods generalize naturally to other blockchain systems,
DAOs, and digital institutions. The concrete research directions that follow are
each motivated by the tensions identified above.

\subsection{Current Challenges in Voltaire Governance}\label{section:voltaire problems}

Before presenting our research directions, we outline key challenges highlighted by Voltaire governance that motivate our research directions that follow.

\paragraph{Centralization risks.}
Governance exhibits concentration along multiple dimensions. Among DReps, the top 10 control 48.7\% of active voting power (2.8B ADA out of 5.8B delegated), and the top 20 control 63.6\%~\cite{cgov-dashboard}; network effects favor incumbents (high-visibility DReps attract delegations, increasing their visibility further). Among SPOs, the Nakamoto coefficient is approximately 20 (meaning 20 pools control $>50\%$ of stake). These figures indicate significant centralizing pressures that could intensify absent counter-mechanisms.

\paragraph{Low and declining participation.}
As of March 2026, approximately 5.8B ADA ($\sim$15.7\% of the $\sim$37B circulating supply) is actively delegated for governance, spread across 76,738 delegators and 551 active DReps (out of 1,544 registered)~\cite{cgov-dashboard}. Moreover, even among active participants, meaningful engagement is sparse: many DReps vote infrequently or rubber-stamp proposals without deep analysis. Low participation creates both a legitimacy deficit (decisions may not reflect true stakeholder preferences) and an attack surface (small coalitions can accumulate disproportionate influence when most stake is dormant).

\paragraph{Stake-is-sticky problem.}
Empirical data reveals striking inertia:\footnote{Based on data retrieved from Cardano DB Sync--\url{https://github.com/IntersectMBO/cardano-db-sync}--a tool that synchronizes on-chain data into a relational PostgreSQL database for analytical querying. We analyzed the complete set of available delegation records in Epochs 507--574 (September 2024 to August 2025). This comprehensive dataset reveals that while 220,784 unique delegators assigned voting power to 1,373 distinct Delegated Representatives (DReps), the vast majority of these delegations remained unchanged throughout the eleven-month observation period. Specifically, 208,160 delegators (94.3\%) maintained their initial DRep choice without ever switching, while only a small fraction (5.7\%) opted to change their representative. These findings suggest that delegation remains largely static, indicating that delegators rarely reassess their initial delegation.} 94\% of delegators have never redelegated their voting power, even when their chosen DRep becomes inactive or ideologically misaligned. This ``delegator-representative commitment bias'' persists despite Cardano's frictionless redelegation (no fees, no lock-up periods). The result is \emph{weight capture}: representatives retain voting power solely due to behavioral inertia, not ongoing alignment or performance. This degrades both representational accuracy and accountability.

\paragraph{Safety-liveness tensions.}
Voltaire's high approval thresholds aim to ensure security: constitutional changes require 67\% DRep + 51\% SPO + Constitutional Committee approval. But these thresholds interact with variable participation in complex ways. When participation is low (say, 20\% of stake active), a 67\% threshold of active stake corresponds to only $\sim$13\% of total stake, which is potentially low enough for an attack. Conversely, when participation is high, the same threshold may create gridlock, as even moderate disagreement blocks amendments. No formal analysis exists of how attack costs, amendment costs, and participation rates interact across Voltaire's parameter space.

\subsection{A Framework for Governance Analysis}\label{section:gg}

To address these challenges systematically, we propose a four-layer architecture for modeling, simulating, and optimizing self-amending governance systems. The architecture--described succinctly in Table~\ref{table:framework}--draws on multiagent systems, mechanism design, and computational social choice, and is directly inspired by ongoing work on executable institutional models~\cite{boellaIntroNORMAS2007,andrighettoDagstuhlBook2013}.

\begin{table}[t]
\centering
\footnotesize
\renewcommand{\arraystretch}{1.2}
\caption{Four-layer architecture for governance analysis.}
\label{table:framework}
\vspace{0.5em}
\begin{tabularx}{\linewidth}{l X}
\hline
\textbf{Layer} & \textbf{Purpose} \\
\hline
Formal Kernel & Executable semantics of collective decision and self-amendment \\
Agent-Based Simulation & Emergent behavior under realistic agent models and attacks \\
Empirical Interface & Calibration from on-chain data; replay of historical decisions \\
Optimization & Multi-objective search over mechanism/parameter space \\
\hline
\end{tabularx}
\end{table}

\subsubsection{Formal kernel}

At the foundation is a minimal, executable model of governance as a state-transition system. The kernel isolates institutional logic from surrounding infrastructure (proposal submission, vote aggregation) and provides a common semantics for heterogeneous mechanisms.

\begin{defbox}[t]
\centering
\caption{Formal kernel for self-amending governance.}
\label{box:kernel}
\vspace{0.5em}

\begin{tcolorbox}[
  colback=gray!5, colframe=gray!60,
  fonttitle=\bfseries, title=Governance Kernel,
  width=0.95\columnwidth, boxrule=0.5pt
]
\small

A governance system is a tuple
$K = (\Bodies, \Types, \Policy, \Effect, \sigma)$, where
$\Bodies$ is a finite set of governance bodies,
$\Types$ is a finite set of proposal types,
$\Policy : \Types \to (\text{Tallies} \to \text{Bool})$ maps each type to a
  decision predicate over aggregate vote tallies,
$\Effect : \Types \to (\text{State} \to \text{State})$ maps each type to a
  deterministic state update, and
$\sigma$ denotes the current governance parameters (thresholds, deposits, etc.).

\medskip
\textbf{Execution.}
Given proposal~$p$ of type~$t$, the environment provides tallies
$\tally_b(p) = (Y_b, N_b, A_b, E_b)$ for each body $b \in \Bodies$.
The kernel evaluates $\Policy(t)\bigl(\{\tally_b(p)\}_{b \in \Bodies}\bigr)$;
if satisfied, it applies $\Effect(t)(\sigma)$ to update state.

\medskip
\textbf{Self-amendment.}
The system is \emph{self-amending} if some $\Effect(t)$ modifies future decision
conditions--e.g., altering $\Policy$ predicates, $\Bodies$ membership, or
parameters in~$\sigma$.

\end{tcolorbox}
\end{defbox}

The kernel enables formal reasoning about governance systems: safety properties (``treasury cannot be fully drained''), expressiveness (``which institutional configurations are reachable?''), and convergence (``does iterative self-amendment stabilize or cycle?''). It also provides a common representation for comparing disparate mechanisms: Voltaire, Tezos, Polkadot, and Optimism can all be encoded in the kernel, enabling apples-to-apples comparison.

\paragraph{Governance as Markov chain optimization under uncertainty.}
From a theoretical standpoint, a governance system induces a Markov chain over states $\sigma \in \Sigma$, where transitions are determined by which proposals emerge and how they are voted upon. The constitution (the tuple $\langle \Bodies, \Types, \Policy, \Effect \rangle$) defines the transition probabilities: given state $\sigma_t$ at time $t$ and environment $\theta_t$ (distribution over proposal types, stakeholder preferences, external events), what is the probability of reaching state $\sigma_{t+1}$?

Constitutional design is thus \emph{optimization under uncertainty}. When enacting rules, we choose a Markov chain whose long-run behavior, including convergence properties, expected welfare, and stability guarantees, remains stable under uncertainty about future environments. Will we face mostly routine parameter updates or existential crises requiring rapid pivots? Will stakeholders remain aligned or fracture into adversarial coalitions? The challenge is that we cannot observe the full distribution $\theta$ in advance, so we must design transition rules that perform well across plausible futures.

This framing unifies our research directions: simulation explores the Markov chain empirically under various $\theta$; safety-liveness analysis characterizes absorbing states and convergence rates; attention-based models ask how agent rationality affects transition probabilities; and parameter optimization searches for constitutional configurations maximizing expected long-run welfare subject to safety constraints.

\paragraph{An example kernel with Voltaire implementation.}

Box~\ref{box:kernel} describes a  model that is a possible governance kernel. This governance kernel is rich enough to describe Voltaire's basic operation, as demonstrated below:
\begin{example}[Voltaire's basic operation as a governance kernel (simplified)]

In the notation of Box~\ref{box:kernel}, Voltaire's basic operation can be described as follows.
\begin{itemize}
\item $\Bodies = \{\text{DReps},\; \text{SPOs},\; \text{CC}\}$
\item $\Types = \{\text{ParamChange},\; \text{Treasury},\; \text{HardFork},\; \text{InfoAction}\}$
\item $\sigma = [\text{drepThreshold} \mapsto 0.67,\; \text{spoThreshold} \mapsto 0.51,\; \ldots]$
\end{itemize}

\noindent Now, for example, for type $t = \text{ParamChange}$, the policy requires $\geq 67\%$ DRep
approval: $\Policy(t)$ checks $Y_{\text{DReps}} / (Y_{\text{DReps}} +
N_{\text{DReps}}) \geq \sigma[\text{drepThreshold}]$, and $\Effect(t)$ updates
the relevant entry in~$\sigma$. 
Also, observe that a proposal changing \texttt{drepThreshold} to
$0.75$ is self-amending: it modifies the $\Policy$ predicate governing future
parameter changes.
\end{example}

\subsubsection{Agent-based simulation}

The kernel specifies institutional rules but says nothing about agent behavior. To study emergent dynamics--how do rational agents delegate? how do coalitions form? how do attacks unfold?--we require agent-based simulation.

We chart a corresponding executable simulation platform implementing the kernel with realistic agent models. Agents are heterogeneous: some vote naively (e.g., based on ideology), others strategically (e.g., vote to maximize expected utility), and still others adapt via learning (e.g., reinforcement learning, LLM-based reasoning). The environment includes:
\begin{itemize}
\item \textbf{Agent types:} delegators (choose a single representative, as per current CIP-1694; experimental variants may model multi-representative delegation), representatives (decide whether/how to vote), attackers (coordinate to manipulate outcomes).
\item \textbf{Behavioral models:} bounded rationality (limited attention), strategic delegation (maximize alignment minus effort cost), commitment bias (inertia in redelegation).
\end{itemize}

This simulation framework supports counterfactual analysis: ``What if DRep approval threshold were 75\% instead of 67\%?'' or ``What if participation dropped to 15\%?'' By replaying governance histories under alternative rules, we can isolate causal effects of design choices.

\subsubsection{Empirical interface}

Models must be grounded in reality. The empirical interface imports on-chain governance data (proposal histories, vote tallies, delegation graphs, treasury movements) and links it to simulation. This enables two workflows:

\begin{itemize}
\item \textbf{Calibration:} Fit agent parameters to match observed behavior. For instance, if 94\% of delegators never redelegate, the ``commitment factor'' in behavioral models should be set accordingly.
\item \textbf{Replay:} Re-run historical decisions under counterfactual rules. Example: Replay Cardano's first 50 ratified proposals under alternative thresholds; measure how many would have passed/failed.
\end{itemize}

Empirical validation is crucial: models that cannot reproduce observed behavior are unlikely to generate reliable predictions. The interface also enables continuous learning: as new governance data accumulates, models are re-calibrated, improving predictive accuracy over time.

\subsubsection{Optimization layer}

With executable models and empirical grounding, optimization becomes concrete. We define governance objectives as functions of simulated trajectories:
\[
\mathsf{Obj}(K, \theta) = (W_1, W_2, \ldots, W_k) \in \mathbb{R}^k
\]
where $K$ is a governance configuration (thresholds, deposits, etc.), $\theta$ represents external environment (participation rates, stake distribution), and $W_i$ are objective dimensions.

Representative objective dimensions include: \textbf{decentralization} (Gini index, Nakamoto coefficient); \textbf{participation rate} (fraction of active stake); \textbf{representational accuracy} (alignment with direct-voting baseline); \textbf{security} (attack cost measured in stake); \textbf{throughput} (ratification latency); and \textbf{stability} (outcome variance and amendment frequency).

The optimization problem is multi-objective:
\[
\max_{K \in \mathcal{K}} \; \mathbb{E}_{\theta \sim \mathcal{D}} [\mathsf{Obj}(K, \theta)]
\]
where $\mathcal{K}$ is the space of feasible configurations and $\mathcal{D}$ is a distribution over environments (e.g., varying participation levels). This yields Pareto frontiers characterizing fundamental tradeoffs: configurations that improve security at cost of throughput, or enhance participation at cost of stability.

Search methods include evolutionary algorithms, Bayesian optimization, and gradient-free optimization over simulation outputs. The result: not a single ``optimal'' governance system, but a menu of configurations achieving different tradeoff points, allowing communities to choose based on their priorities.

\subsection{Research Directions}\label{section:research directions}

We now outline specific research directions enabled by this framework, each addressing one or more challenges identified above.

\subsubsection{Simulation for Voltaire governance}

\textbf{Goal:} Implement the Voltaire kernel as a simulation framework, supporting diverse agent models, attack scenarios, and counterfactual analyses.

Core requirements are:
\begin{itemize}

\item Faithful implementation of Voltaire's rules (all seven action types, all approval thresholds, all effects).

\item Agent heterogeneity: naive voters, strategic voters, learning agents (RL, LLM-augmented).

\item Attack scenarios: vote buying, bribery, coordination attacks, Sybil attacks (via stake concentration).

\item Empirical calibration: 
import delegation graphs, voting histories, stake distributions from Cardano mainnet.

\item Extensibility: plugin architecture for new mechanisms (participatory budgeting, retroactive funding, etc.).

\end{itemize}

\paragraph{Open questions include:}
\begin{itemize}

\item \emph{Agent sophistication:} How to model bounded rationality? Agents cannot evaluate all proposals deeply; which heuristics do they use? How do sophisticated agents (e.g., DReps with expert advisors) differ from naive ones? 

\item \emph{Calibration from sparse data:} On-chain data reveals votes and delegations but not internal reasoning (Why did this DRep vote Yes? Why did this delegator choose that representative?). How to infer latent parameters (attention budgets, ideological positions) from observable actions?

\item \emph{Attack prioritization:} Which attack vectors are most urgent to simulate? Vote buying? Flash loan attacks on governance tokens? Misinformation campaigns to manipulate delegation?

\end{itemize}

The deliverable is a Cardano-specific simulation framework, providing a reference implementation for Voltaire, documentation, and example scenarios. It should enable reproducible governance research and community-driven improvement.

\subsubsection{Safety-liveness analysis with dynamic thresholds}

\textbf{Goal:} Characterize how static approval thresholds interact with variable participation to create security-liveness tradeoffs, and explore adaptive threshold mechanisms.
Voltaire's static thresholds (e.g., 67\% of active DRep stake) provide weak security when participation is low and create gridlock when participation is high. There is still no formal analysis of how attack costs and amendment costs scale with participation.

\paragraph{Open questions include:}
\begin{itemize}
\item \emph{Attack cost vs. participation:} How does the cost of passing a malicious proposal (measured in ADA required) scale with participation rate? Is there a critical participation threshold below which attacks become feasible?
\item \emph{Amendment cost vs. participation:} How does the difficulty of passing legitimate proposals scale with participation? At what level do high thresholds create effective veto power for small coalitions?
\item \emph{Dynamic thresholds:} Should thresholds adapt to participation? Can adaptive mechanisms be gamed by suppressing participation?
\item \emph{Formal guarantees:} Can we prove properties like ``Attack always costs $\geq X$ ADA'' or ``Legitimate proposal with $\geq Y\%$ support passes within $Z$ epochs''?
\end{itemize}
Methods include game-theoretic equilibrium analysis (modeling attackers and defenders strategically), simulation (exploring parameter spaces empirically), and formal verification (mechanized proofs).

\subsubsection{Attention-based social choice and incentive design}

\textbf{Goal:} Understand how limited agent attention affects voting quality and design mechanisms to incentivize meaningful (not just nominal) participation.
Agents have limited attention, and current Voltaire provides no DRep compensation. This leads to chronic under-provision of informed engagement.
A basic model of this may be as an attention-effort game where each agent $i$ chooses effort level $e_i \geq 0$ at cost $c(e_i)$, which increases voting accuracy: $\Pr(\text{vote correctly} \mid e_i) = 0.5 + \gamma \cdot e_i$. Agents trade off collective benefit (correct decision) against individual cost, typically yielding under-provision equilibria.

\paragraph{Open questions include:}
\begin{itemize}
\item \emph{Equilibrium characterization:} For symmetric agents, what is the unique equilibrium? How does it depend on effectiveness $\gamma$, population size $n$, and stake distribution?
\item \emph{Welfare loss:} How far below optimal is equilibrium effort? Can we bound the price of anarchy?
\item \emph{Multiple proposals:} With limited attention budgets across $m$ proposals, how do agents allocate effort? Which equilibrium concepts apply?
\end{itemize}
Methods include game-theoretic analysis (Nash equilibria for symmetric case, mixed strategies generally), simulation (complex cost structures), and mechanism design (optimal payment rules). The outcome informs DRep compensation: rather than paying per vote (rewarding quantity), pay for quality proxies like alignment with expert consensus or ex-post accuracy.

\subsubsection{Delegation dynamics and commitment bias}

\textbf{Goal:} Quantify how behavioral inertia in delegations affects decentralization and representational accuracy, and design interventions to mitigate harmful commitment.
Empirical inertia appears to be high: many Cardano delegators may never redelegate, even when representatives become inactive or misaligned. This creates weight capture, where representatives retain power due to stickiness rather than performance.
A basic model is via a commitment factor $s_d \in [0,1]$ governing re-delegation probability. When a better representative emerges with alignment gain $\Delta c$, switching probability is $P_{\text{switch}} = (1 - s_d) \cdot (\Delta c)^{s_d}$. High $s_d$ implies strong commitment; low $s_d$ implies frictionless switching.

\paragraph{Open questions include:}
\begin{itemize}
\item \emph{Decentralization vs. accuracy tradeoff:} How does commitment affect voting weight concentration (Gini, Nakamoto) versus representational accuracy (alignment with direct-voting baseline)?
\item \emph{Weight capture dynamics:} In oscillating drift scenarios (opinions shift left-right-left), can we characterize when representatives accumulate and retain delegations purely via inertia?
\item \emph{Optimal friction:} Is there an optimal commitment level balancing stability (too low causes churn) against alignment (too high causes capture)?
\item \emph{Interventions:} Which mechanisms reduce harmful commitment: mandatory re-delegation periods, nudges when representatives become inactive, or time-decay of delegation strength?
\item \emph{Strategic representatives:} How do strategic opinion adjustments (representatives shift to retain delegations) affect dynamics?
\end{itemize}
Methods include agent-based simulation (varying commitment factor, measuring outcomes), behavioral modeling (calibrating from empirical data), and counterfactual analysis (simulating interventions). This informs design of DRep accountability mechanisms and optimal re-delegation friction.

\subsubsection{Multi-objective parameter optimization}

\textbf{Goal:} Characterize Pareto-optimal configurations across Voltaire's $\sim$30 governable parameters to balance competing objectives.
Joint parameter setting remains difficult. Approval thresholds, deposits, quorums, and term lengths interact in complex ways, and no principled method exists for choosing them.
Representative objective dimensions include: \textit{decentralization} (Gini index, Nakamoto coefficient); \textit{participation rate} (fraction of active stake); \textit{representational accuracy} (alignment with direct-voting baseline); \textit{security} (attack cost in stake); \textit{throughput} (ratification latency); and \textit{stability} (outcome variance, amendment frequency).

\paragraph{Open questions include:}
\begin{itemize}
\item \emph{Pareto frontier:} What are Pareto-optimal configurations? Can we characterize fundamental tradeoffs (e.g., high security + high throughput vs. high participation + high accuracy)?
\item \emph{Sensitivity analysis:} Which parameters matter most? Can we rank parameters by influence on outcomes?
\item \emph{Robustness:} Do good configurations remain good under changing environments (e.g., participation dropping from 30\% to 20\%)?
\item \emph{Constraint satisfaction:} How to search parameter space subject to hard safety constraints (e.g., ``treasury never fully drains'')?
\end{itemize}
Methods include simulation-based optimization (evolutionary algorithms, Bayesian optimization), multiobjective search, and sensitivity analysis. Deliverables include Pareto frontier visualizations, parameter sensitivity rankings, and recommended configurations for different community priorities (security-focused vs. agility-focused).

\subsubsection{Constitutional guardrails and meta-governance}

\textbf{Goal:} Design constraints on self-amendment that permit adaptation while preserving critical invariants and preventing capture.
Voltaire is self-amending: governance actions can modify their own approval conditions. This enables adaptability but also risks instability or capture via iterative rule changes.

\paragraph{Open questions include:}
\begin{itemize}
\item \emph{Automated vs. human guardrails:} Which constraints can be enforced by on-chain scripts (e.g., ``cumulative treasury withdrawals must not exceed the Net Change Limit'') versus requiring human judgment (e.g., ``proposal aligns with ecosystem goals'')?
\item \emph{Meta-governance design:} Should rules governing rule changes be harder to modify than operational rules? Does this create useful stability or harmful rigidity?
\item \emph{Formal verification:} Can we prove safety properties about self-amending systems (e.g., ``treasury balance always $> 0$'' or ``at least one body can always propose amendments'')?
\item \emph{Amendment cycles:} Can iterative self-amendment lead to cycles or does it converge? Are cycles harmful (instability) or benign (democratic oscillation)?
\end{itemize}

Methods include formal verification, mechanism design (constitutional constraints preserving invariants while permitting adaptation), and simulation (exploring reachability of governance state space; in particular, simulations of the dynamics).

\subsubsection{Adversarial proposal timing}

\textbf{Goal:} Characterize how rational actors exploit predictable participation cycles by strategically timing proposal submission, and design countermeasures.

Participation in Voltaire governance is not uniform over time. Predictable low-participation windows (e.g., holidays, weekends, periods following high-proposal congestion) create exploitable attack surfaces. A proposer can time submission to coincide with low DRep activity, increasing the probability that their proposal passes with a small active coalition.

A formal instantiation: let $P_t$ be participation (active stake fraction) at epoch $t$. An adversary with stake $w_a$ submits proposal $p$ at time $t^*$ chosen to minimize the coalition size needed to ratify:
\[
t^* = \arg\min_t \; \frac{\theta \cdot P_t \cdot S}{w_a}
\]
where $S$ is total stake and $\theta$ is the applicable DRep threshold. Lower $P_t$ means a smaller absolute stake suffices.

\paragraph{Open questions include:}
\begin{itemize}
\item \emph{Empirical patterns:} Are there measurable participation troughs in Cardano governance data? Do they correlate with calendar patterns or with proposal congestion?
\item \emph{Attack cost:} How much cheaper is a successful attack during a trough vs.\ average participation? Is the difference practically significant for treasury withdrawals?
\item \emph{Countermeasures:} Can submission windows, mandatory notice periods, or participation-adaptive delays mitigate timing attacks without harming liveness for legitimate proposers?
\item \emph{Detection:} Can anomalous timing patterns (proposals submitted at predictable troughs) be flagged on-chain as a warning signal to delegators?
\end{itemize}

Methods include empirical analysis (participation time-series from mainnet data), game-theoretic modeling (adversary maximizing attack success probability subject to participation dynamics), and mechanism design (submission window rules).

\subsubsection{Emergency governance and legitimate overrides}

\textbf{Goal:} Formalize the interface between Voltaire's deliberative governance and time-critical emergency intervention, leveraging the Legitimate Override framework~\cite{elemtalmon2025overrides} for optimizing the design, taking into account the negative effects of centralization.

Voltaire's mechanisms assume well-functioning governance: proposals proceed through multi-epoch voting with deliberative thresholds. Crises (exploits, oracle failures, market shocks) unfold on timescales of minutes to hours, far faster than any governance vote can conclude. When governance is too slow, many protocols rely on emergency mechanisms (pauses, freezes, emergency councils) that operate outside normal governance channels. Elem and Talmon~\cite{elemtalmon2025overrides} develop a complementary framework for these scenarios: bounded, time-limited overrides with explicit scope constraints and constitutional-style safeguards.

\paragraph{Open questions include:}
\begin{itemize}
\item \emph{Delegation to emergency bodies:} Under what conditions should Voltaire delegate authority to an Emergency Committee, and how should those delegated powers be scoped, time-bounded, and revocable?
\item \emph{Standing centralization cost:} How does the existence of emergency override capability affect Voltaire's perceived decentralization and trust model, even when never exercised?
\item \emph{Scope--authority matching:} The Scope $\times$ Authority taxonomy~\cite{elemtalmon2025overrides} suggests that different threat types require different intervention precision levels. Can Voltaire's parameter governance be extended to include emergency mechanism selection?
\item \emph{Post-crisis accountability:} How should emergency actions be ratified, reviewed, or reversed through normal Voltaire governance after the crisis passes?
\end{itemize}

This connects normal governance (Voltaire) with crisis governance (Legitimate Overrides), filling a gap in the current specification where emergency response is left to off-chain coordination.

\section{A Call to Action}\label{sec:conclusion}

We have provided a \textbf{complete specification} of Voltaire governance
(Section~\ref{sec:specification}), establishing it as a precisely-defined object
of study: a live, self-amending, multi-stakeholder governance system with
${\sim}30$ tunable parameters, formal on-chain execution, and a rich empirical
record of proposals, votes, and delegations. Building on this, we introduced a
four-layer architecture--formal modeling, agent-based simulation, empirical
calibration, and multi-objective optimization--as a methodological foundation
for blockchain governance research (Section~\ref{section:gg}). Finally, we
presented a \textbf{concrete research agenda} grounded in real challenges facing Voltaire
(Section~\ref{sec:researchagenda}).

\textbf{The stakes} are not just academic: Voltaire governs a network securing
billions of dollars in assets; its design choices--thresholds, deposits, term
lengths, quorum rules--directly affect who can influence protocol evolution, how
much attacks cost, and whether governance remains legitimate under adversarial
conditions. Getting these choices wrong has cascading consequences; getting them
right requires the kind of rigorous, multi-method approach we have outlined.

Our research program is admittedly ambitious, but its scope reflects the
complexity of decentralized governance: no single method suffices, and progress
demands integrating formal theory, simulation, empirical analysis, and
optimization. Voltaire, we argue, serves as an ideal laboratory--a live system
with high parameterizability and rich on-chain data to ground and validate our
models.
More broadly, our vision is one of \textbf{governance as an engineering
discipline}: a field equipped with rigorous theory, validated models, reusable
tools, and empirical benchmarks. Just as distributed systems research combines
theory (consensus protocols, impossibility results) with practice (building and
stress-testing systems at scale), governance research should combine mechanism
design with empirical validation and iterative improvement. This vision extends
well beyond blockchains. DAOs, open-source projects, and increasingly collectives
of AI agents all require governance mechanisms that balance autonomy with
coordination, stability with adaptability, and security with efficiency. The
framework developed here is designed to generalize to these domains as well.

We close with a \textbf{call to action}. The governance research community can accelerate
progress by investing in shared infrastructure: a simulation framework
implementing the Voltaire kernel for reproducible counterfactual analysis;
standardized, machine-readable governance datasets spanning Cardano, Tezos,
Polkadot, and major DAOs; benchmark attack and governance scenarios enabling
apples-to-apples comparison of mechanism designs; and formal, machine-readable
encodings of governance rules supporting automated verification. Such
infrastructure would allow the field to move from ad hoc design to principled
engineering. Governance engineering is nascent, and we invite the community to
join us in building it.

\bibliographystyle{plain}
\bibliography{bib}
\end{document}